\documentstyle[sprocl]{article}

\def\be{\begin{equation}}
\def\ee{\end{equation}}
\def\bea{\begin{eqnarray}}
\def\eea{\end{eqnarray}}

\begin{document}

\title{SPINOR DESCRIPTION OF A GENERAL SPIN-$J$ SYSTEM}

\author{V. R. VIEIRA, P. D. SACRAMENTO}

\address{ Centro de F\'{\i}sica das Interac\c{c}\~{o}es Fundamentais,
Instituto Superior T\'{e}cnico, \\ Av. Rovisco Pais, 
1096 Lisboa Codex, Portugal} 

\maketitle\abstracts{
We consider a spin coherent states description of a general quantum
spin system. It is shown that it is possible to use the spin-$1/2$
representation to study the general spin-$J$ case. We identify 
the $1/2$ spinor components as the homogeneous coordinates of the
projective space associated to the complex variable that labels
the coherent states and establish a relation between the two-component
spinor and the bosonic Schwinger representation of a spin operator.
We rewrite the equations of motion, obtained from the path integral for 
the evolution operator or partition function, in terms of the 
$1/2$ spinor and define the effective Hamiltonian of its evolution.
}

\section{Introduction}
Coherent states provide a convenient way to represent quantum operators in 
terms of c-number functions. They have been used extensively in the literature in
various fields like quantum optics, condensed matter and quantum field theory~\cite{1}.
They are closest to a classical description in the sense that they minimize the
Heisenberg uncertainty relations. In particular, spin coherent states have proved
useful to deal with spin systems~\cite{2}. Since the algebra of the spin operators
is more involved than that for boson or fermion operators, the construction of the
spin coherent states is somewhat more complicated. Recently, we have pursued the
similarities to the boson coherent states and have shown that it is possible
to establish a closer parallelelism than previously realized. In particular, we
obtained generalized representatives for spin operators~\cite{3,4} extending results
obtained previously for bosons~\cite{5}. Also, we considered the path integral for the
transition amplitude and the partition function~\cite{6} and considered quantum Monte
Carlo algorithms using coherent states~\cite{7}.

The handling of spin operators is traditionally a difficult problem implying, in
general, unphysical states that have to be projected out. There are methods which do
not include extra states~\cite{8,9} but they are not easily generalized to spin values 
greater than one. Another advantage of the spin coherent states is that the spin 
value $J$ enters only as a parameter and therefore all spin values can be handled 
similarly. 
The price to pay is a description in terms of continuous variables instead of a discrete
number of states. The spin coherent states are therefore a convenient basis to perform
a semi-classical expansion for large values of the spin~\cite{10} and they appear 
naturally in the quantization of a semi-classical spin theory~\cite{11,12}. 

The fact that $J$ is only a parameter in the formalism suggests that one can 
reformulate the spin-$J$ case in terms of the $J=1/2$ case. In this paper we 
emphasize that indeed most results obtained for a general $J$-value are simply 
related to the same expression for $J=1/2$ which is in general simpler to obtain.

\section{Spinor versus stereographic projection descriptions} 

The spin coherent states can be defined by~\cite{2}
\be
|\alpha> = R(\theta,\varphi) |JJ> = \frac{1}{(1+|\alpha|^{2})^{J}} 
e^{\alpha \hat{J}_{-}} |JJ>,
\ee
where $R(\theta, \varphi)$ is a rotation through an angle $\theta$ about 
the axis ${\vec n}=(\sin \varphi, -\cos \varphi,0)$, normal to the $z$-axis and 
to the vector 
${\vec m}=(\sin \theta \cos \varphi, \sin \theta \sin \varphi, \cos \theta)$,
and the state $|JJ>$ is the top extreme eigenstate of $\hat{J}^{2}$ and
$\hat{J}_{z}$.
The variable $\alpha = \tan{\frac{\theta}{2}} e^{i \varphi}$ defines the stereographic
projection from the south pole of the sphere to the plane passing through the 
equator, with complex coordinates $\alpha, \alpha^*$.
Under this projection, distances in the sphere are mapped into the plane 
according to 
$\left(d\vec{s}\right)^{2}=d\theta^{2}+\sin^{2}\theta d\varphi^{2} 
=4d|\alpha|^{2}/(1+|\alpha|^{2})^{2}$,
and areas or spherical angles according to
$d\Omega=\sin\theta d\theta d\varphi 
= 4d^{2}\alpha/(1+|\alpha|^{2})^{2}$,
where $d^{2}\alpha=d\Re\alpha d\Im\alpha$.
Under the rotation considered above the vectors of the spherical basis 
are rotated according to $(0,0,1) \rightarrow \vec{m}$ and
$1/2(1,\pm i,0) \rightarrow \vec{h}_{\mp}$, where 
$2(1+\alpha^* \alpha) \vec{h}_{-}=(1-\alpha^2, i(1+\alpha^2), -2\alpha)$ and
$\vec{h}_{+}=\vec{h}_{-}^*$.
These vectors satisfy $\vec{m}^{2}=1$, 
$\vec{h}_{-}\cdot\vec{h}_{+}=\frac{1}{2}$, 
$\vec{h}_{\pm}^{2}=0$ and $\vec{m}\cdot\vec{h}_{\pm}=0$. 
The coherent states are not orthogonal and their overlap is 
\bea
<\alpha^{\prime}|\alpha> &=& \frac{(1+\alpha^{\prime *}\alpha)^{2J}}
{(1+|\alpha^{\prime}|^{2})^{J} (1+|\alpha|^{2})^{J}} \\
&=& \left[ e^{i \Phi} \frac{(1+\cos{\Theta})}{2} \right]^{J}.
\eea
where $\Phi$ is the area of the spherical triangle with vertices $\vec{m}$,  
$\vec{m}^{\prime}$ and $\vec{e}_{z}$ and $\Theta$ is the angle between the two 
directions defined by $\vec{m}$ and $\vec{m}^{\prime}$.
The coherent state $|\alpha>$ depends on the variables $\alpha$ and
$\alpha^{*}$, which should be considered as independent. 
It is convenient to define matrix elements depending
only on two complex variables (and not four)~\cite{3}. The same happens in the
case of bosons where the so-called holomorphic representation is used~\cite{13,14}. 
The easiest way to use this type of representation is to define
new states~\cite{3}
$\|\alpha> = e^{\alpha \hat{J}_{-}} |JJ>$
which depend only on $\alpha$. Their overlap is
$<\beta \|\hat{1}\|\alpha > = (1+\beta^{*}\alpha)^{2J}$
and the decomposition of the identity becomes
\be
\frac{2J+1}{\pi} \int d^{2}\alpha 
\frac{\|\alpha><\alpha\|}{(1+|\alpha|^{2})^{2(J+1)}} = \hat{1}.
\ee
The coherent states $\|\alpha>$ and $<\alpha\|$ for 
spin $\frac{1}{2}$ are given by the spinors
\bea
\Lambda(\alpha) & = & \left( \begin{array}{c} 1 \\ \alpha \end{array} \right) \\
\Lambda^{\dagger}(\alpha) & = & ( \begin{array}{cc} 1 & \alpha^{*} \end{array} ),
\eea
respectively, and 
one can write
\bea
\Lambda^{\dagger}(\alpha)\Lambda(\alpha) & = & 1+\alpha^{*}\alpha \\
<\alpha\|\hat{1}\|\alpha>& = &\left[ \Lambda^{\dagger}(\alpha)\Lambda(\alpha)\right]^{2J}.
\eea

The matrix element of an operator $\hat{F}$ is defined as 
$F(\beta^{*},\alpha) = <\beta\|\hat{F}\|\alpha>$.
It depends only on $\alpha $ and $\beta^{*}$, and it does not depend on 
$\alpha^{*}$ or $\beta$. 
One can then use $\alpha^{*}$ instead of $\beta^{*}$, and treat 
$\alpha$ and $\alpha^{*}$ as independent variables, without any loss of generality. 
For some operators like the density matrix it is convenient to use the 
so-called diagonal
representative given by the weight function of the decomposition of the
operator in a superposition of coherent states projectors~\cite{5}.
In connection with the holomorphic representation it is convenient to
define the diagonal representative
$\bar{f}(\alpha,\alpha^{*})$ given by~\cite{3}
\be
\hat{F} = \frac{2J+1}{\pi} \int d^{2}\alpha \|\alpha>
\bar{f}(\alpha,\alpha^{*}) <\alpha\|.
\ee

Since the spin coherent states are not eigenvectors of the angular momentum operators,
it is important to know their action on them.
These relations were obtained before~\cite{4}
\bea
e^{\beta \hat{J}_{-}}\|\alpha> & = & \|\alpha+\beta> \nonumber \\
e^{\beta^{*} \hat{J}_{+}}\|\alpha> & = & (1+\beta^{*}\alpha)^{2J} \|\frac{\alpha}
{1+\beta^{*}\alpha}> \nonumber \\
e^{\beta \hat{J}_{z}}\|\alpha> & = & e^{\beta J}\|e^{-\beta}\alpha>.
\eea
The action of the infinitesimal operators on the coherent states follows
immediately~\cite{4} and can be summarized as 
\be
\hat{\vec{J}}\|\alpha>=\{J\vec{m}+\vec{h}_{-}[(1+\alpha^{*}\alpha)
\frac{\partial}{\partial\alpha}-2J\alpha^{*}]\}\|\alpha>
\ee
The matrix elements of the operators $\hat{J}_{\pm}$ 
and $\hat{J}_{z}$ are given by~\cite{1,2,3,4}
$<\alpha|\hat{\vec{J}}|\alpha> = J \vec{m}$.
Similarly, the diagonal representatives are given by 
$\vec{j}(\alpha,\alpha^*) = (J+1) \vec{m}$.
From eq. (11) follows that the coherent states minimize the uncertainty relation 
for any two spin components since, from the Schwarz inequality, in order that a
state $|\phi >$
minimizes the uncertainty relation, the states $(J_i-<J_i>) |\phi>$, with $i=\pm, 0$, 
must be linearly dependent. 

The operators $\hat{D}(g)$ of the spin-$J$ representation of the rotation group 
can be parameterized in different manners. The most intuitive one is to define 
the axis of rotation, defined by the unit vector $\vec{m}$ and the angle of 
rotation $\phi$. We have then
$\hat{D}(g)=e^{i\vec{\phi}\cdot\hat{\vec{J}}}$,
where $\vec{\phi}=\phi\vec{m}$. A specially useful set of parameters can be 
found looking at the form of these operators for $J=1/2$, where they reduce to
\bea
\hat{{\cal D}}(g) & = & e^{i\vec{\phi}\cdot\frac{\vec{\sigma}}{2}} \nonumber \\
 & = & \cos\frac{\phi}{2} + i\frac{\vec{\phi}}{\phi} \cdot 
\vec{\sigma}\sin\frac{\phi}{2} \nonumber \\
 & = & \left( \begin{array} {cc} \mu & -\nu^{*} \\ \nu & \mu^{*} \end{array} \right),
\eea
where
$\mu=\cos\frac{\phi}{2}+i\cos\theta\sin\frac{\phi}{2}$ and 
$\nu=i \sin\theta e^{i\varphi}\sin\frac{\phi}{2}$.
These two complex numbers are not independent, since they satisfy
$|\mu|^{2}+|\nu|^{2}=1$.
The inverse of an operator is obtained replacing $\mu$ by $\mu^{*}$ and 
$\nu$ by $-\nu$. Using these parameters, the disentangling theorem~\cite{1} is written as
\bea
\hat{D}(g) & = & e^{-\frac{\nu^{*}}{\mu^{*}}\hat{J}_{+}} e^{-2 \ln \mu^{*} 
\hat{J}_{z}} e^{\frac{\nu}{\mu^{*}} \hat{J}_{-}} \\
& = & e^{\frac{\nu}{\mu}\hat{J}_{-}} e^{2 \ln \mu \hat{J}_{z}} 
e^{-\frac{\nu^{*}}{\mu} \hat{J}_{+}}
\eea

The action of a general rotation on a coherent state is~\cite{4}
\be
\hat{D}(g) \|\alpha> = (\mu-\nu^{*}\alpha)^{2J} \|\frac{\nu+
\mu^{*}\alpha}{\mu-\nu^{*}\alpha}>.
\ee
The matrix element of the operator $\hat{D}(g)$ is then given by
\be
<\alpha^{\prime}\|\hat{D}(g)\|\alpha> = [\mu+\alpha^{\prime *}\nu-\nu^{*}\alpha+
\mu^{*}\alpha^{\prime *}\alpha]^{2J}
\ee
The matrix element of the spin $1/2$ operator $\hat{{\cal D}}(g)$ 
between these spin $\frac{1}{2}$ coherent states is given by
\bea
{\cal D}(g;\alpha^{*},\alpha) & = & 
\mu+\alpha^{*}\nu-\nu^{*}\alpha+\mu^{*}\alpha^{*}\alpha \nonumber \\
& = & \Lambda^{\dagger}(\alpha)
\hat{{\cal D}}(g)\Lambda(\alpha).
\eea
One concludes then that the matrix element for general spin $J$ is simply
\be
<\alpha\|\hat{D}(g)\|\alpha> = [{\cal D}(g;\alpha^{*},\alpha)]^{2J}
\ee
i.e. the power $2J$ of the corresponding spin $\frac{1}{2}$ matrix element,
with eq. (8) being a special case. 
These results are easily understood if one uses the tensorial product of $2J$ 
spins $\frac{1}{2}$, $\vec{\hat{J}}_{a}$, $a=1,\ldots,2J$, and write 
$|JJ>=|\frac{1}{2} \frac{1}{2}>\cdots|\frac{1}{2} \frac{1}{2}>$ and 
$\vec{\phi}.\hat{\vec{J}}= \sum_{a=1}^{2J} \vec{\phi}.\hat{\vec{J}}_{a}$. 
The matrix elements of a general rotation between coherent states factorize 
then in the product of the factors for each spin $\frac{1}{2}$, where the 
spin $\frac{1}{2}$ coherent states are given by eqs. (5,6). 

The diagonal representative of $\hat{D}(g)$ is
\be
\bar{d}(g;\alpha,\alpha^{*}) = \frac{1}{[\mu^{*}-\alpha^{*}\nu + 
\nu^{*}\alpha +\mu\alpha^{*}\alpha ]^{2(J+1)}}.
\ee
which can be rewritten as
\be
\bar{d}(g;\alpha,\alpha^{*}) = \frac{1}{[{\cal D}(g^{-1};\alpha^{*},\alpha)]^{2(J+1)}}.
\ee
i.e. the diagonal representative of $\hat{D}(g)$, for general spin $J$, is 
given by the power $-2(J+1)$ of the matrix element, for spin $\frac{1}{2}$, 
of the operator $\hat{D}^{-1}(g)$. 

A direct way to calculate the matrix elements of products of powers of spin 
operators consists in differentiating the appropriate generating function 
$<\alpha\ \| e^{u \hat{J}_{a}} e^{v \hat{J}_{b}}\cdots \|\alpha>$ and using 
the reduction of this matrix element to the $2J$ power of the corresponding 
spin $\frac{1}{2}$ matrix element. 
Repeated use of the identity 
$\sigma_{a} \sigma_{b} = \delta_{ab } +i\epsilon_{abc} \sigma_{c}$, 
where $\sigma_{a}$ are the Pauli matrices, reduces the calculation to 
$m_{a}=<\alpha\|\sigma_{a}\|\alpha>/<\alpha\|1\|\alpha>$. 
One concludes then that any correlation function is a function only 
of the vector $\vec{J}=J\vec{m}$. The same applies to diagonal representatives.
In practice, however, correlation functions and cumulants, in particular, are 
most easily evaluated using the fact that the insertion at right (or at left) 
of additional spin operators into matrix elements or diagonal representatives 
of some operator is given by the differential operators 
$J\vec{m}+\vec{h}_{-}(1+\alpha^{*}\alpha)\frac{\partial}{\partial\alpha}$ and
$(J+1)\vec{m}-\vec{h}_{+}(1+\alpha^{*}\alpha)\frac{\partial}{\partial\alpha^{*}}$
(or their complex conjugates), acting on those matrix elements or diagonal 
representatives~\cite{15}.

If we consider the action of a rotation $\hat{{\cal D}}(g)$ on a general 
$\frac{1}{2}$ spinor 
\be
\Lambda_{\lambda}(\alpha) = \left( \begin{array}{c} \lambda \\ \alpha \end{array} \right)
\ee
we find that the relation between the components of the rotated 
$\Lambda_{\lambda}^{\prime}(\alpha)=\hat{{\cal D}}(g)\Lambda_{\lambda}(\alpha)$ and original spinor 
$\Lambda_{\lambda}(\alpha)$ 
can be written as
\be
\frac{\alpha^{\prime}}{\lambda^{\prime}}=
\frac{\nu+\mu^{*}\frac{\alpha}{\lambda}}{\mu-\nu^{*}\frac{\alpha}{\lambda}}.
\ee
From eq.(15) it follows that the variable $\alpha$, labeling the spin 
coherent states, has the same transformation law as the ratio $\alpha/\lambda$, of the two components of the spinor.
This is the relation, found in geometry, between an affine space and the 
projective space associated with it.
More precisely, we identify the $1/2$ spinor components as the homogeneous 
coordinates of the projective space associated to the complex variable which 
labels the coherent states.
This coordinate transforms according to a bilinear or M\"obius transformation. 
One recovers the well known isomorphism between the $SU(2)$ group and the group 
of the bilinear transformations.
The variable $\lambda$ is the rescaling necessary to have the first component 
of the spinor equal to one. 

Under a transformation in which the spinors
$\Lambda_{\lambda}(\alpha)$ and  
$\Lambda_{\lambda}^{\dagger}(\alpha)$  
transform according to
$\Lambda_{\lambda}(\alpha^{\prime})=U\Lambda_{\lambda}(\alpha)$ and 
$\Lambda_{\lambda}^{\dagger}(\alpha^{\prime})=\Lambda_{\lambda}^{\dagger}(\alpha
) V$,
where $U$ and $V$ are not necessarily related, the Jacobian for the change of 
variables is given by
\be
(|\lambda^{\prime}|^{2})^{2} d^{2}\frac{\alpha^{\prime}}{\lambda^{\prime}} = 
\det U \det V (|\lambda|^{2})^{2} d^{2}\frac{\alpha}{\lambda}.
\ee
The two components of the spinors transform separately as the numerator and the 
denominator of the bilinear transformations. 
This is very useful in practical calculations. 
For example, in the calculation of the integrals involving matrix elements 
or diagonal representatives, one can replace matrix elements between 
coherent states by matrix elements between spinors, and use rotations and not 
bilinear transformations, to simplify their evaluation. 
This, combined with the relation between the spin $J$ and spin $1/2$ 
quantities, allows a great simplification of the calculations. 

Also, the connection between rotations of vectors and rotations 
of spinors expressing the fact that the spin is a vector operator and given 
by ${\cal R}(\vec{m})\cdot\vec{\sigma} = 
\hat{\cal D}\vec{m}\cdot\vec{\sigma}\hat{\cal D}^{-1}$ 
can be immediately obtained using the identity
\be
\vec{m}(\alpha)\cdot\vec{\sigma} = \frac{2 \Lambda \Lambda^{\dagger}}{
\Lambda^{\dagger}\Lambda} -\hat{1}
\ee
in terms of the spinors $\Lambda$ and $\Lambda^{\dagger}$.
Due to the homogeneity of the expression for $\vec{m}$ the vector $\vec{J}=
J \left(\Lambda_{\lambda}^{\dagger} \vec{\sigma} \Lambda_{\lambda}\right)/
\left(\Lambda_{\lambda}^{\dagger}\Lambda_{\lambda}\right)$ 
coincides with the vector $\vec{J}=J \vec{m}$ previously defined,
when we make the replacement $\alpha \rightarrow \alpha/\lambda$, $\alpha^* \rightarrow
\alpha^*/\lambda^*$. 

\section{Path integral and equations of motion}

The transition amplitude
\be
\bar{U} (\alpha_{f}^{*},t_{f};\alpha_{i},t_{i}) = <\alpha_{f}|| T_{t}
e^{-\frac{i}{\hbar} \int_{t_{i}}^{t_{f}} dt \hat{{\cal H}}(t)}
||\alpha_{i}>
\ee
can be obtained by the path integral expression~\cite{6}
\be
\bar{U} (\alpha_{f}^{*},t_{f};\alpha_{i},t_{i}) = \int {\cal
D}^{2}\mu (1+\alpha_{f}^{*} \alpha(t_{f}) )^{J} e^{\frac{i}{\hbar}
\int_{t_{i}}^{t_{f}} dt {\cal L}} (1+\alpha^{*}(t_{i})
\alpha_{i})^{J}
\ee
where
\be
{\cal L} = \frac{\hbar}{i} \frac{J}{1+\alpha^{*}\alpha} (
\frac{d\alpha^{*}}{dt} \alpha - \alpha^{*} \frac{d\alpha}{dt}) -
{\cal H}_{c}(\alpha^{*}, \alpha,t)
\ee
is the Lagrangian, in which ${\cal H}_{c}(\alpha^{*}, \alpha,t)$ is
the normalized matrix element of the Hamiltonian. 
The dependence on $\alpha$ and $\alpha^{*}$ is through $\vec{m}$ only, as explained above.
The two boundary terms appear naturally from the construction of the path 
integral~\cite{6} and are necessary to have the boundary conditions 
$\alpha(t_{i}) = \alpha_{i}$ and $\alpha^{*}(t_{f}) =\alpha_{f}^{*}$, when varying the action. 
Note that $\alpha(t_{f})$ is not the complex conjugate of $\alpha_{f}^{*}$ 
and that $\alpha^{*}(t_{i})$ is not the complex conjugate of $\alpha_{i}$. 
They are the values taken by the variables $\alpha(t)$ and $\alpha^{*}(t)$, 
satisfying these boundary conditions, at the times $t=t_{f}$ and $t=t_{i}$, 
respectively. 
In particular, in the saddle-point approximation they are obtained
solving the classical equations of motion for $\alpha(t)$ and
$\alpha^{*}(t)$ and evaluating their values at $t_{f}$ and $t_{i}$,
respectively. 
These terms have been shown to be necessary to obtain the exact solution using 
the saddle-point approximation in the case of a spin in a time-independent 
magnetic field, for instance~\cite{6}. 
In the discrete approximation the integration measure is
$ {\cal D}^{2}\mu= \Pi^{N-1}_{k=1} \frac{2J+1}{\pi}
\frac{d^{2}\alpha_{k}}{(1+\alpha^{*}_{k}\alpha_{k})^{2}}$. 
It can be seen as a consequence of the quantization of a classical 
problem with constraints~\cite{12}. 

The usual transition amplitude is therefore given by
\be
<\alpha_{f} | T_{t} e^{- \frac{i}{\hbar} \int_{t_{i}}^{t_{f}} dt
\hat{{\cal H}}(t)} |\alpha_{i}> = \int {\cal D}^{2}\mu \left(
\frac{1+ \alpha_{f}^{*}\alpha(t_{f})}{1+\alpha_{f}^{*}\alpha_{f}}
\right)^{J} e^{\frac{i}{\hbar} \int_{t_{i}}^{t_{f}} dt {\cal L}}
\left( \frac{1+ \alpha^{*}(t_{i}) \alpha_{i}}{1+
\alpha_{i}^{*}\alpha_{i}} \right)^{J}
\ee
where $\alpha_{f}$ and $\alpha_{i}^{*}$ are the complex conjugates of
$\alpha_{f}^{*}$ and $\alpha_{i}$, respectively. 
The path integral representation for the diagonal representative of the 
evolution operator is similar~\cite{6}.

It can be checked that, both for matrix elements and diagonal representatives, 
the classical equations of motion, obtained varying the Lagrangian term and 
the boundary condition terms, are the classical limit of the Dyson-Schwinger 
equations, with the appropriate boundary conditions~\cite{6}. 
Also, the Poisson brackets, or more precisely, the Dirac brackets for the 
quantization of a theory with constraints, of the matrix elements or diagonal 
representatives of any two spin components satisfy the commutator algebra of 
the quantum spin operators.

The path integral expression for the partition function can be easily
obtained and one finds using matrix elements
\be
Z= \int {\cal D}^{2} \mu (t) e^{-\frac{1}{\hbar}
\int_{0}^{\beta \hbar} d\tau [-\hbar \frac{J}{1+\alpha^{*}\alpha}
(\frac{d\alpha^{*}}{d\tau}\alpha - \alpha^{*} \frac{d\alpha}{d\tau})
+ {\cal H}_{c} (\alpha^{*},\alpha) ]}
\ee
with periodic boundary conditions in the imaginary time. The expression for 
the case of the diagonal representative is similar~\cite{6}.

Geometrically, the free term in the Lagrangian is the area in the sphere 
defined by the trajectory under consideration and by two great arcs from the 
north pole to the initial and final points of the trajectory.
It can be rewritten as
\be
{\cal L}_{0}=\frac{d\vec{J}}{dt}\times\vec{J}\cdot\vec{\varphi},
\ee
where $\vec{{\varphi}}=\frac{\vec{u}}{J+\vec{J}\cdot\vec{u}}$, with 
$\vec{u}=\vec{e_{z}}$, reflecting the choice of the $z$ axis as the 
quantization axis in the construction of the coherent states. 
In general, $\vec{u}$ can be any arbitrary unit vector. 
Changing it, the free Lagrangian is modified by 
$\delta{\cal L}_{0}=\frac{d}{dt}
\left(\frac{J}{J+\vec{J}\cdot\vec{u}}\vec{J}\times\vec{u}\cdot\delta\vec{u}\right)
$
i.e., the total derivative of a function of the initial and final positions 
(the difference in the areas defined by each of these points and the old 
and new $\vec{u}$).
Conversely, under a rotation, i.e., under a change of variables of the form 
$\alpha^{\prime}=
\frac{\nu+\mu^{*}\alpha}{\mu-\nu^{*}\alpha}$, the free Lagrangian changes by 
${\cal L}_{0}^{\prime}-{\cal L}_{0}=\frac{d }{dt}
\ln(\frac{\mu^{*}-\alpha^{*}\nu}{\mu-\nu^{*}\alpha}) $
which is also the total derivative of a function of the initial and final 
positions.
Finally, under a spin inversion, in which $\vec{J}\rightarrow -\vec{J}$, 
$\alpha\rightarrow -\frac{1}{\alpha^{*}}$, 
$\alpha^{*}\rightarrow -\frac{1}{\alpha}$, 
the free Lagrangian transforms according to 
${\cal L}_{0}\rightarrow -{\cal L}_{0}+\frac{d}{dt}(-iJ\ln(\frac{\alpha^{*}}{\alpha}))=
-{\cal L}_{0}-2J\frac{d\varphi}{dt}$, with the geometrical interpretation that 
the sum of the areas defined by a closed trajectory encircling the poles and 
viewed from them is $4\pi$, the total area of the sphere.

The Lagrangian is totally rewritten in terms of the vector $\vec{J}$. The
integration on the sphere can also be written as $1/J^2 d^3J \delta (\vec{J}^2-
J^2)$. We arrive then at the path integral for spins
\be
\int {\cal D} \vec{J} \delta (\vec{J}^2-J^2) e^{i \int dt {\cal L}}
\ee

Similarly, a formulation involving only the spinors can also be found.
Making the rescaling $\alpha \rightarrow \alpha/\lambda$ and
$\alpha^* \rightarrow \alpha^*/\lambda^*$, the integration measure becomes
$d \Omega = |\lambda|^2 d^2\alpha / (\Lambda^{\dagger} \Lambda)^2$. In this
expression the values of $\lambda$ and $\lambda^*$ are arbitrary but fixed.
In order to integrate over $d^2\lambda$ one introduces two delta functions
$\delta (\Re \lambda -\lambda_r)$ and 
$\delta (\Im \lambda-\lambda_i)$. Using the identity
\be
\delta (f(x,y)) \delta(g(x,y)) \left|\frac{\partial(f,g)}{\partial(x,y)}\right|_{x=x_0, 
y=y_0} =
\delta (x-x_0) \delta (y-y_0)
\ee
where $x_0$ and $y_0$ are such that $f(x_0,y_0)=g(x_0,y_0)=0$, one is able to
give other forms to the path integral. In particular defining
\be
A = \left( \begin{array}{c} a_1 \\ a_2 \end{array} \right) = 
i \sqrt{2J} 
\frac{r(\lambda^*,\lambda)}{\sqrt{1+\left(\frac{\alpha}{\lambda}\right)^*
\frac{\alpha}{\lambda}}} 
\left( \begin{array}{c} 1 \\ \frac{\alpha}{\lambda} \end{array} \right) \\
\ee
where $r(\lambda^*, \lambda)$ is an arbitrary but not constant function, one finds
immediately the path integral expression found by Jevicki and Papanicolaou~\cite{11}
\be
Z= \int \prod_{i=1}^{2} {\cal D}^2 a_i \prod_t \delta [A^{\dagger}(t) A(t) -2J]
\delta [\phi(t)] \{\phi,N\}_{PB} e^{i \int dt {\cal L}}
\ee
derived using the Schwinger representation for spins and using the 
Dirac-Faddeev theory of constrained systems, extended to the case of  
second class constraints~\cite{12}, to make the restriction to the 
physical space $A^{\dagger} A= N=2J$ and where $\phi=a_1+a_1^*$ 
is the gauge fixing condition. In terms of the
function $r(\lambda^*,\lambda)$ this restriction and the gauge condition 
become $|r|^2=1$ and $r^*=r$, respectively.
 
The classical equations of motion can be obtained varying the Lagrangian 
term and the boundary condition terms. In the case of the path integral 
for the matrix element, we find
\bea
i\hbar \frac{2J}{(1+\alpha^{*}\alpha)^{2}} \frac{d\alpha}{dt} & = &
\frac{\partial {\cal H}}{\partial \alpha^{*}} \nonumber \\ -i\hbar
\frac{2J}{(1+\alpha^{*}\alpha)^{2}} \frac{d\alpha^{*}}{dt} & = &
\frac{\partial {\cal H}}{\partial \alpha}
\eea
with $\alpha(t_{i}) = \alpha_{i}$ and
$\alpha^{*}(t_{f}) = \alpha_{f}^{*} $ as boundary conditions, 
as expected~\cite{16}.
Using \mbox{$(1+\alpha^* \alpha) \partial \vec{m}/\partial \alpha=2 \vec{h}_+$}
and its complex conjugate~\cite{17},
the equations of motion can be rewritten as
\bea
i\hbar\frac{2}{(1+\alpha^{*}\alpha)^{2}} \frac{d\alpha}{dt} & = &
\frac{\partial {\cal H}}{\partial\vec{J}}\cdot
\frac{2h_{-}}{1+\alpha^{*}\alpha} \nonumber \\ -i\hbar
\frac{2}{(1+\alpha^{*}\alpha)^{2}}\frac{d\alpha^{*}}{dt} & = &
\frac{\partial {\cal H}}{\partial\vec{J}}\cdot
\frac{2h_{+}}{1+\alpha^{*}\alpha}
\eea
where $\vec{J}=J\vec{m}$. In these equations 
$\vec{H}^{eff}=\frac{\partial{\cal H}}{\partial\vec{J}}$ is the effective 
magnetic field acting on the spin. Making the rescaling 
$\alpha\rightarrow\frac{\alpha}{\lambda}$ and 
$\alpha^{*}\rightarrow\frac{\alpha^{*}}{\lambda^{*}}$ and defining the 
effective Hamiltonian 
$\tilde{{\cal K}}=\vec{H}^{eff}\cdot\frac{\vec{\sigma}}{2}$, 
acting on the 
spinors, one finds that these equations of motion are equivalent 
to  
\be
i\hbar\frac{d\Lambda_{\lambda}}{dt}=
\tilde{{\cal K}}\Lambda_{\lambda}
\ee
and its hermitian conjugate.
Since $\tilde{{\cal K}}$ is an hermitian operator, the norm of the spinors, 
i.e. 
$\Lambda_{\lambda}^{\dagger}\Lambda_{\lambda}=
\lambda^{*}\lambda+\alpha^{*}\alpha$, 
is conserved.
The equation of motion of the moment $\vec{J}$, 
i.e., of the 
expectation value of the Pauli matrices in the spinor, 
is
\be
\frac{d\vec{J}}{dt}=\frac{\partial {\cal H}}{\partial \vec{J}}\times \vec{J}
\ee
which is the equation of precession 
in the 
effective field $\frac{\partial {\cal H}}{\partial \vec{J}}$.

The equation of motion of the spinor can be formally solved by
\be
\Lambda_{\lambda}(\alpha)=\tilde{U}(t,t_{0})\Lambda_{\lambda}^{0}(\alpha)
\ee
where 
\be
\tilde{U}(t,t_{0})=T_{t} e^{-\frac{i}{\hbar}\int_{t_{0}}^{t}dt\tilde{K}}
\ee
is the evolution operator corresponding to the effective Hamiltonian 
$\tilde{K}$. 
One can take advantage of the connection between vector rotations and 
spinor rotations by defining the operator in spinor space
\be
\tilde{{\cal J}}=\vec{J}\cdot\vec{\sigma}
\ee
having the property $\tilde{J}^2=J^2$. The equation of precession becomes
\be
i\hbar\frac{d\tilde{{\cal J}}}{dt}=[\tilde{{\cal K}},\tilde{{\cal J}}]
\ee
and can be solved by
\be
\tilde{{\cal J}}=\tilde{U}\tilde{{\cal J}}(0)\tilde{U}^{-1}
\ee

The case of many spins can be easily treated, repeating this procedure for 
each spin. The path integral follows immediately, with the coupling between 
spins arising from the Hamiltonian ${\cal H}$. The saddle point equations 
follow from there and one finds the mean field approximation in which each 
spin i precesses in the effective field 
$\vec{H}_{i}^{eff}=\frac{\partial {\cal H}}{\partial \vec{J}_{i}}$
resulting, in particular, from the interaction with external fields and the 
other spins.
One introduces Pauli matrices $\vec{\sigma}_{i}$ for each spin. The effective
Hamiltonian for each spin is 
$\tilde{{\cal K}_{i}}=\vec{H}^{eff}_{i}\cdot\frac{\vec{\sigma}_{i}}{2}$
The total effective Hamiltonian is 
$\tilde{{\cal K}}=\sum_{i}\tilde{{\cal K}_{i}}$. 
It is the sum of single spin terms, reflecting the character of the mean field 
approximation in which each spin precesses in the mean produced by the 
others. 
We note that in the treatment of continuous spin chains the matrix $\tilde{J}$ is the
central quantity for the Lax representation used by Takhtajan~\cite{18} and Jevicki and
Papanicolaou~\cite{11} in the derivation of nontrivial solutions of the soliton or instanton
type.

\section{Summary} 

We have replaced a description of a quantum spin operator in terms of
classical variables (defined by the stereographic projection of the sphere from its 
south pole to the equatorial plane) by a spinor description (two-component spinor)
and replaced the action of a rotation on the spin coherent states (which involves
bilinear transformations of the complex variables) by the action of a rotation
(matrices operating in spinor space) on the two-component spinor. 
This reformulation appears naturally for $J=1/2$ but we have shown that it generalizes 
in a straightforward way to all spin values. 
We have noted that this spinor description is related
to the description of a spin system by Schwinger bosons. 
Both descriptions enlarge the space and a projection to the proper subspace is 
required. 
We have reexpressed the equations of motion in terms of the $1/2$ spinor and 
obtained simple expressions that have a formal appearance familiar from standard 
dynamics of two-level systems.
This enables a simpler treatment of a general spin-$J$ system. Interacting spin
systems are dealt with more conveniently in terms of an evolution
operator that acts on spinor space and in terms of a transfer matrix that connects
different points in real space. Also, we have made contact to previous treatments
where the Lax representation is used in the context of the search for nontrivial
solutions of the saddle-point equations via the inverse scattering method.


\begin{thebibliography}{99}
\bibitem{15} The vectors $\vec{h}_{+}, \vec{h}_{-}$ 
are the eigenvectors of the projectors 
$P_{ab}^{+} = \frac{1}{2} (\delta_{ab}^{\perp} + i \epsilon _{abc} m_{c})=
2h_{+}^{a}h_{-}^{b}$ and $P_{ab}^{-}=\left(P_{ab}^{+}\right)^{*}$, 
where $\delta_{ab}^{\perp}=\delta_{ab}-\delta_{ab}^{||}$ 
and $\delta_{ab}^{||}=m_{a}m_{b}$ are the transversal and longitudinal 
projectors relative to the vector $\vec{m}$. See ref. [4].
\bibitem{16} In some applications, involving circles or straight lines as 
trajectories, the mapping of circles and straight lines into circles and 
straight lines by the bilinear transformations becomes useful.
The parameterization of the solutions and handling of the boundary conditions 
becomes easier using the invariance of the cross ratio of four points
under the bilinear transformations, or in a circle, under complex conjugation. 
The same can be done using the spinors in projective space.
\bibitem{17} The variations of $\vec{h}_{\pm}$ with $\alpha$ and $\alpha^*$ are
given by $(1+\alpha^* \alpha) \partial \vec{h}_{-}/\partial \alpha = -\vec{m} +
\alpha^* \vec{h}_{-}$ and $(1+\alpha^* \alpha) \partial \vec{h}_{+}/\partial \alpha =
-\alpha \vec{h}_{+}$ and their complex conjugates.
\end{thebibliography}
\end{document}